\documentstyle[eqsecnum,aps]{revtex}

\begin{document}
\draft


\title{Superheating fields of superconductors: Asymptotic analysis and
numerical results}

\author{Andrew J. Dolgert\cite{ajdemail}, S.\ John Di Bartolo\cite{sjdemail},
and Alan T.\ Dorsey\cite{atdemail}}

\address{Department of Physics,  University of Virginia, McCormick Road,
Charlottesville, Virginia~~22901}

\date{\today}

\maketitle

\begin{abstract}

The superheated Meissner state in type-I superconductors is studied both
analytically and numerically within the framework of Ginzburg-Landau
theory.  Using the method of matched asymptotic expansions we
have developed a systematic expansion for the solutions of the Ginzburg-Landau
equations in the limit of small $\kappa$, and have determined the maximum
superheating field $H_{\rm sh}$ for the existence of the metastable,
superheated Meissner state as an expansion in powers of $\kappa^{1/2}$.  Our
numerical solutions of these equations agree quite well with the
asymptotic solutions for $\kappa<0.5$.  The same asymptotic methods are also
used to study the stability of the solutions, as well as a modified version
of the Ginzburg-Landau equations which incorporates nonlocal
electrodynamics.  Finally, we compare our numerical results for the
superheating field for large-$\kappa$ against recent asymptotic results
for large-$\kappa$, and again find a close agreement.
Our results demonstrate the efficacy of the method of matched
asymptotic expansions for dealing with problems in inhomogeneous
superconductivity involving boundary layers.

\end{abstract}


\section{Introduction}

In equilibrium,  the superconducting Meissner phase of a bulk type-I
superconducting sample exists below a thermodynamic critical field $H_c$
(or below $H_{c1}$ in a type-II superconductor);  above this field the
sample reverts to the normal phase (or the flux lattice phase in a
type-II superconductor).  However, because the phase transition
in both cases is first-order it is possible to superheat the Meissner
phase and delay the transition to fields well above $H_c$ or $H_{c1}$.
This superheated, metastable Meissner phase is
eventually destroyed at a maximum superheating field $H_{\rm sh}$.
Understanding the origin and stability of the superheated state is the
first step in providing a complete description of the time-dependent
collapse of the Meissner phase, which is important for many applications
of type-I superconductors.  For instance, there have been recent
proposals to use superheated type-I superconductors as detectors
for elementary particles, so that the sample acts as a superconducting
``bubble chamber'' \cite{detectors}.   The passage of a sufficiently energetic
particle through the sample would initiate the transition to the
normal state.   Measuring the superheating
field also provides one of the few methods of experimentally
determining the Ginzburg-Landau parameter $\kappa$ in type-I
superconductors \cite{burger}.

The precise value of the maximum superheating field may depend upon
extrinsic factors such as defects in the sample and  sample preparation
and geometry.  If these effects can be minimized
then the limit of superheating is determined by the boundaries and geometry
of the sample.  The simplest
and most widely studied geometry is a superconducting half-space with
a magnetic field applied parallel to the surface of the superconductor,
which is the geometry considered in the remainder
of this paper.  To model the superconductor we will use the
Ginzburg-Landau (GL) equations, which provide an accurate description of
the surface behavior provided the coherence length $\xi$ is large compared
to microscopic length scales \cite{degennes}.
Previous studies of superheating in type-I superconductors have used a variety
of heuristic methods to determine the behavior of the GL equations
near the surface. Ginzburg \cite{ginzburg58} inferred the leading
$\kappa^{-1/2}$ dependence of the superheating field from the form of the
Ginzburg-Landau equations.  Apparently unaware of
Ginzburg's work, the Orsay group \cite{orsay} used a variational argument
to show that $H_{\rm sh}/H_c \approx 2^{-1/4} \kappa^{-1/2}$.
By combining an ingenious guess for the behavior of the superconducting
order parameter near the surface with a variational calculation,
Parr \cite{parr76} was able to calculate the next order correction to
the Orsay group's result.  In addition to this analytical work there
has also been a great deal of numerical
work on solving the GL equations in small-$\kappa$ limit
\cite{matricon67,kramer68,fink69,bolley}, which is reviewed in
Ref.~\cite{fink}.  The numerical results appear to confirm at least
some of the analytical work, although admittedly over a somewhat
restricted range of $\kappa$.  One deficiency common to all of the
previous analytical approaches is an {\it ad-hoc} construction
of approximate solutions of the GL equations, leaving us without a
procedure for systematically improving upon these approximations.
In addition, the issue of the {\it stability} of the solutions in
the small-$\kappa$ limit seems not to have been addressed
rigorously (for one attempt see \cite{galaiko68}).

In this paper we re-examine the problem of superheating in type-I
superconductors by using the method of matched asymptotic expansions
\cite{bender,vandyke} to solve the GL equations in the small-$\kappa$ limit.
This method was originally developed to treat
boundary layer problems in fluid mechanics \cite{vandyke} in a controlled
and systematic fashion, and
is particularly well suited to the superheating problem, as all of
the technical difficulties arise due to a ``boundary layer'' at the surface.
Using this method we can calculate the superheating
field in the small-$\kappa$ limit as an asymptotic expansion in
powers of $\kappa^{1/2}$, construct
uniform asymptotic expansions (i.e., expansions valid for all $x$ as
$\kappa\rightarrow 0$) for the order parameter and magnetic field,
determine the stability of the solutions, and treat
nonlocal electrodynamic effects.  Where appropriate we compare our
asymptotic results against numerical solutions of the GL equations,
and we generally find excellent agreement.
We have chosen to present our results in detail, as the methods are probably
unfamiliar to most physicists.
Matched asymptotic expansions have recently been used by Chapman
\cite{chapman95} to study superheating in the large-$\kappa$ limit, and our
results complement his work. This paper, taken together with Chapman's work,
demonstrates that these perturbation
methods can provide a powerful calculational tool for solving problems
in inhomogeneous or nonequilibrium superconductivity.

The remainder of this paper is organized as follows.  In Sec.~II we describe
our numerical methods for solving the GL equations.  In Sec.~III we
develop the method of matched asymptotic expansions for the solution
of the GL equations in the small-$\kappa$ limit, and determine the
first three terms in the expansion of $H_{\rm sh}$ in powers of
$\kappa^{1/2}$.  In addition, we construct uniform expansions for the
order parameter and magnetic field and compare them against our
asymptotic results.  In Sec.~IV we examine the second variation of the
GL free energy, $\delta^2 {\cal F}$, in order to determine the stability
of our solutions.  This is done for both one and two-dimensional
perturbations.  In Sec.~V the method is generalized to treat
nonlocal electrodynamics.  Section~VII compares Chapman's asymptotic
expansion for $H_{\rm sh}$ for large-$\kappa$ with our numerical results,
and we find remarkably good agreement.  Finally, Sec.~VIII is a summary
and discussion of our results.

\section{Numerical methods}

The GL free energy of a superconducting sample occupying the
half-space $x>0$ is
\begin{equation}
{\cal F}[f,q] = \int_{x>0}\:d^3r\left[\frac{1}{\kappa^2}({\bf\nabla} f)^2+
    \frac{1}{2}(1-f^2)^2+f^2 {\bf q}^2+({\bf H}_a-{\bf\nabla\times q})^2\right]
\label{energy}
\end{equation}
where $\kappa$ is the GL parameter, $f$ is the amplitude of
the superconducting order parameter,  ${\bf q}$ is the gauge-invariant
vector potential (${\bf h}={\bf\nabla\times q}$), and ${\bf H}_a$ is the
applied magnetic field.  The lengths are in units
of the penetration depth $\lambda$ and fields are in units of $\sqrt{2}H_c$.
Minimizing this expression with respect to both $f$ and ${\bf q}$ results
in the GL equations.  In one dimension, with $f=f(x)$ and
${\bf q}=(0,q(x),0)$, these equations are
\begin{equation}
{1\over \kappa^2} f'' - q^2 f + f - f^3 = 0,
\label{GL1}
\end{equation}
\begin{equation}
q'' - f^2 q = 0 ,
\label{GL2}
\end{equation}
\begin{equation}
 h = q'.
\label{GL3}
\end{equation}
The task at hand is to solve these equations numerically for a
superconducting half-space and to find the largest possible applied field
($H_{\rm sh}$) which permits a superconducting solution.  To
insure that no current passes through the boundary at $x=0$ and that
the sample is totally superconducting infinitely far from the surface,
we impose the boundary conditions
\begin{equation}
f'(0) = 0,  \qquad f(x)\rightarrow 1\ \ {\rm as} \ x\rightarrow \infty.
\label{BC1}
\end{equation}
Since the field at the surface must equal the applied field $H_a$, and the
field infinitely far from the surface must equal 0, we impose the boundary
conditions
\begin{equation}
h(0) = H_a, \qquad q(x) \rightarrow 0 \ \ {\rm as} \ x\rightarrow \infty.
\label{BC2}
\end{equation}

For $\kappa\rightarrow0$, we rescale the equations as $x'=\kappa x$ making
the new unit of length the correlation length~$\xi$.  Since $\xi\gg\lambda$
in this limit, a numerical solution over a domain much larger than $\xi$
would insure that the regions of rapid change for $f$ and $h$ would be
included.  (For small $\kappa$, we find that solving for $x'<500$ is
sufficient.)  In the large $\kappa$ limit, we use the rescaled equations
again, but we increase the size of the domain depending on the value of
$\kappa$.  (The equations must be solved for domains as large as $x'<10^4$
for values of $\kappa\sim10^3$.)

The equations can be solved using the relaxation method \cite{press}.
By replacing these ordinary differential equations with  finite
difference equations, one can start with a guess to the solution and
iterate using a multi-dimensional Newton's method until it relaxes
to the true solution.  In order to more accurately pick up the detail near
the boundary, we choose a grid of discrete points with a higher density
near $x=0$.  In particular we choose a density which roughly varies as
the inverse of the distance from the boundary.  (For low $\kappa$ our
density, in units of mesh points per coherence length, varies approximately
from $10^7$ near the boundary to $10^3$ at the farthest point from the
boundary, while for high $\kappa$ it varies from $10^5$ to $10^{-2}$.)

$H_{\rm sh}$ can be found in the following way.  For a given value of
$\kappa$ an initial guess is made where there is no applied field and the
sample is completely superconducting ($f\equiv1, q\equiv0, h\equiv0$).
The field $H_a$ is then ``turned up'' in small increments.  For each value
of $H_a$ a solution is sought using the result from the previous lower
field solution as an initial guess.  Eventually a maximum value for $H_a$
is reached, above which one of two things happens: our algorithm fails to
converge to a solution or it converges to the normal (nonsuperconducting
solution).  This maximum value of $H_a$ is the numerical result for
$H_{\rm sh}$.
Using this algorithm, $H_{\rm sh}(\kappa)$ can be found for a wide range of
$\kappa$'s.  Each run (for a given $\kappa$) takes about 60 cpu minutes
on an IBM RS 6000/370. We find it sufficient for the purposes of this paper
to deal with superheating field values for $10^{-3}<\kappa<10^3$.

\section{Asymptotic expansions for small-$\kappa$}

In this section we will develop an asymptotic expansion for the superheating
field for small-$\kappa$, using the method of matched asymptotic
expansions \cite{bender,vandyke}.
For small-$\kappa$ the dominant length scale is the coherence length $\xi$,
so it is natural to have $\xi$ serve as our unit of length.  This is
achieved by rescaling $x$ by $\kappa$, introducing a new dimensionless
coordinate $x'=\kappa x$.  The resulting GL equations in these
``outer variables'' are
\begin{equation}
f'' - q^2 f + f - f^3 = 0,
\label{GL4}
\end{equation}
\begin{equation}
\kappa^2 q'' - f^2 q = 0 ,
\label{GL5}
\end{equation}
\begin{equation}
 h = \kappa q',
\label{GL6}
\end{equation}
with the primes now denoting differentiations with respect to $x'$.

{\it Outer solution.}  In order to obtain the outer solutions expand
$f$, $q$, and $h$ in powers of $\kappa$:
\begin{equation}
f = f_0 + \kappa f_1 + \kappa^2 f_2 + \ldots,
\label{expand1}
\end{equation}
\begin{equation}
q = q_0 + \kappa q_1 + \kappa^2 q_2 + \ldots,
\label{expand2}
\end{equation}
\begin{equation}
h = h_0 + \kappa h_1 + \kappa^2 h_2 + \ldots.
\label{expand3}
\end{equation}
Substituting into Eqs.~(\ref{GL4})-(\ref{GL6}), at $O(1)$ we have
\begin{equation}
f_0'' - q_0^2 f_0 + f_0 - f_0^3=0,
\label{outer1}
\end{equation}
\begin{equation}
 - f_0^2 q_0 = 0.
\label{outer2}
\end{equation}
Since we want $f\rightarrow 1$ as $x'\rightarrow \infty$, the only possible
solution to Eq.~(\ref{outer2}) is $q_0 = 0$.  We can then immediately
integrate Eq.~(\ref{outer1}),
\begin{equation}
f_0(x') = \tanh \left( {x' + x_0}\over \sqrt{2} \right),
\label{outer3}
\end{equation}
with $x_0$ an arbitrary constant.  To $O(\kappa)$, the outer
equations are
\begin{equation}
f_1'' - 2q_0f_0q_1 - q_0^2f_1 + f_1 - 3f_0^2 f_1 = 0,
\label{outer4}
\end{equation}
\begin{equation}
 - f_0^2 q_1 - 2 f_0q_0 f_1 = 0,
\label{outer5}
\end{equation}
\begin{equation}
h_0 = 0.
\label{outer6}
\end{equation}
Once again, the only solution to Eq.~(\ref{outer5}) is $q_1 = 0$;
substituting this into Eq.~(\ref{outer4}), we find $f_1 =C_1 f_0'$,
with $C_1$ a constant:
\begin{equation}
f_1 = {C_1\over \sqrt{2}} {\rm sech}^2\left( { x'+x_0 \over \sqrt{2}} \right).
\label{outer7}
\end{equation}
We can continue in this manner; at every order $q_n= 0$, $h_n=0$,
and $f_n = C_n f^{(n)}_0$, with the $C_n$'s constants which are determined
by matching onto the inner solution.

{\it Inner solution.}  The outer solution breaks down within a boundary
layer of $O(\kappa)$ near the surface.  This suggests introducing a rescaled
inner coordinate $X=x'/\kappa$, so that $X=O(1)$ within the boundary layer.
It is also possible to rescale $f$ and $q$, with the hope that this
will lead to a tractable inner problem.  Such a rescaling must lead to
a successful matching of the inner and outer solutions; i.e., the inner
solutions as $X\rightarrow \infty$ must match onto the outer solutions as
$x'\rightarrow 0$.  Since $f_0(0) = \tanh (x_0/\sqrt{2})$, then assuming
that $x_0\neq0$ we have $f_0(0)= O(1)$, indicating
that the order parameter should not be rescaled in the inner region;
therefore we set $f(x') = F(X)$  in the inner region. However, from the
outer solution for the vector potential we see that the only constraint
on $q(X)$ in the inner region is that $q(X)\rightarrow 0 $ as
$X\rightarrow\infty$ (presumably exponentially).  Therefore, we
are free to rescale $q$ by $\kappa$ in the inner region, hopefully in
a way which simplifies the inner equations.  One possibility is
$q(x')=\kappa^{-\alpha} Q(X)$; substituting this into the GL equations,
Eqs.~(\ref{GL4})-(\ref{GL6}), we see that unless $2\alpha$ is an
integer, fractional powers of $\kappa$ will be introduced into the inner
equations, contradicting our expansion of $f$ and $q$ in integer powers of
$\kappa$ in the outer region.  Therefore, the simplest assumption is that
$\alpha=1/2$, leading to the following choice for the inner variables:
\begin{equation}
x'= \kappa X,\quad f(x') = F(X),\quad q(x') = \kappa^{-1/2} Q(X), \quad
h(x') = H(X).
\label{inner1}
\end{equation}
In these variables Eqs.~(\ref{GL4})-(\ref{GL6}) become
\begin{equation}
F'' - \kappa Q^2 F + \kappa^2 (F - F^3) = 0,
\label{inner2}
\end{equation}
\begin{equation}
Q'' - F^2 Q = 0,
\label{inner3}
\end{equation}
\begin{equation}
\kappa^{1/2} H =  Q',
\label{inner4}
\end{equation}
where now the primes denote differentiation with respect to $X$.
The boundary conditions are
\begin{equation}
F'(0)=0, \qquad H(0) = H_a.
\label{inner4a}
\end{equation}

The next step is to expand the inner solutions in powers of $\kappa$:
\begin{equation}
F = F_0 + \kappa F_1 + \kappa^2 F_2 + \ldots,
\label{inner5}
\end{equation}
\begin{equation}
Q= Q_0 + \kappa Q_1 + \kappa^2 Q_2 + \ldots,
\label{inner6}
\end{equation}
\begin{equation}
H = \kappa^{-1/2} H_0 +  \kappa^{1/2} H_1 + \ldots.
\label{inner7}
\end{equation}
Note that there is no term of $O(1)$ in the expansion for $H$, since
we would be unable to match such a term to the outer solution.
Using the boundary condition $H(0)=H_a$ leads to
\begin{equation}
H_a = \kappa^{-1/2} H_0(0) + \kappa^{1/2} H_1(0) + \ \ldots.
\label{inner7a}
\end{equation}
Substituting these expansions into Eqs.~(\ref{inner2})-(\ref{inner4}),
at $O(1)$ we obtain
\begin{equation}
F_0'' = 0, \qquad Q_0'' - F_0^2 Q_0 = 0, \qquad H_0 = Q_0'.
\label{inner9}
\end{equation}
Solving these equations subject to the boundary conditions (\ref{inner4a})
(we also need $Q_0\rightarrow 0 $ as $x\rightarrow \infty$ in order to
match onto the outer solution), we obtain
\begin{equation}
F_{0}(X)   = A_0, \qquad Q_0(X) = B_0 e^{-A_0 X},
                  \qquad H_0(0) = -A_0 B_0,
\label{inner10}
\end{equation}
with $A_0$ and $B_0$ constants.  In what follows we will assign $F_n(0) = A_n$
and $Q_n(0)=B_n$ for notational simplicity.  The $O(\kappa)$ equations are
\begin{equation}
F_1'' = Q_0^2 F_0, \qquad Q_1'' - F_0^2 Q_1 = 2 F_0 Q_0 F_1,
\qquad H_1 = Q_1'.
\label{inner11}
\end{equation}
Solving with the boundary condition $F_{1}'(0)=0$, we obtain
\begin{equation}
F_1(X) = A_1 + { B_0^2 \over 4 A_0}\left[ 2 A_0 X +   e^{-2 A_0 X}
                       - 1\right],
\label{inner12}
\end{equation}
\begin{eqnarray}
 Q_1(X) &=& e^{-A_0 X} \left\{ B_1 - {B_0^3\over 16 A_0^2} \biggl[
            1 - e^{-2A_0 X} \right.  \nonumber \\
     & & \qquad  \left. + 16 {A_0^2 A_1 \over B_0^2} X
          + 4 A_0^2 X^2 \biggr] \right\},
\label{inner13}
\end{eqnarray}
\begin{equation}
  H_1(0)= -{1\over 8} {B_0^3 \over A_0} - A_0B_1 - A_1 B_0.
\label{inner13a}
\end{equation}
Finally, to $O(\kappa^2)$ we have for $F_2$
\begin{equation}
F_2'' = - F_0 + F_0^3 + 2Q_0Q_1F_0 + Q_0^2F_1,
\label{next1}
\end{equation}
the solution of which (with $F_2'(0) = 0$) is
\begin{eqnarray}
F_2(X) &=& {17 \over 128}{B_0^4\over A_0^3} + {1\over 4} {B_0^2 A_1 \over
A_0^2}
      - {1\over 2} {B_0 B_1 \over A_0} + A_2  + \left( B_0B_1 - {3\over 32}
      {B_0^4 \over A_0^2} \right) X - {1\over 2} A_0(1-A_0^2) X^2 \nonumber \\
 & & \quad + \left[ {1\over2} {B_0 B_1 \over A_0}
  - {1\over 4}{B_0^2 A_1\over A_0^2}- {5\over 32}{B_0^4\over A_0^3}
  - \left( {1\over 8}{B_0^4\over A_0^2} +{1\over 2} {B_0^2 A_1\over
A_0}\right)X
     - {1\over 8} {B_0^4\over A_0} X^2\right] e^{-2A_0 X} \nonumber \\
 & & \qquad + {3\over 128} {B_0^4 \over A_0^3} e^{-4A_0 X}.
\label{inner14}
\end{eqnarray}
The expression for $Q_2$ is even more unwieldy, and is not needed in
what follows.

{\it Matching.}  To determine the various integration constants which have
been introduced we must match the inner solution to the outer solution.
Since the outer solution for $q$ is simply $q=0$, and all of our
inner solutions decay exponentially for large $X$, the matching is
automatically satisfied for $q$, as well as for the magnetic field $h$.
To match the inner and outer solutions
for the order parameter, we are guided by the {\it van Dyke matching principle}
\cite{vandyke}, which states that the $m$ term inner expansion of the
$n$ term outer solution should match onto the $n$ term outer expansion of the
$m$ term inner solution.  In our case we will take $m=3$ and $n=2$.
Therefore, write the two term outer solution
$f_0(x') + \kappa f_1(x')$ in terms of the inner
variable $X$, and expand for small $\kappa$, keeping the first three terms in
the expansion in powers of $\kappa$:
\begin{eqnarray}
f_0(\kappa X) + \kappa f_1(\kappa X) & \sim&
       \tanh \left({x_0\over \sqrt{2}}\right)
        + \kappa\,  {\rm sech}^2 \left({x_0 \over \sqrt{2}}\right)
              {1\over \sqrt{2}}\left[ C_1 + X \right] \nonumber \\
  & & \qquad  + \kappa^2 {\rm sech}^2 \left({x_0 \over \sqrt{2}}\right)
      \tanh \left({x_0\over \sqrt{2}}\right)\left[- C_1 X
         - {X^2\over 2}\right].
\label{match1}
\end{eqnarray}
Next, write the three term inner solution
$F_0(X) + \kappa F_1(X)+\kappa^2 F_2(X)$ in terms
of the outer variable $x'$, and expand for small $\kappa$, this time keeping
the first two terms of the expansion:
\begin{eqnarray}
F_0(x'/\kappa) + \kappa F_1(x'/\kappa) + \kappa^2 F_2(x'/\kappa) & \sim&
  A_0 + {B_0^2 \over 2} x'  - {1\over 2} A_0(1-A_0^2) x'^2 \nonumber \\
& & \qquad+ \kappa \left[ A_1-{B_0^2 \over 4 A_0} + \left( B_0 B_1 - {3\over
32}
       {B_0^4 \over A_0^2}\right)x'  \right].
\label{match2}
\end{eqnarray}
By writing both expressions in terms of $x'$, and equating the various
coefficients of $x'$ and $\kappa$, we see that the expansions do indeed
match if we choose
\begin{equation}
A_0 = \tanh\left({x_0\over \sqrt{2}}\right),
\label{match3}
\end{equation}
\begin{equation}
\qquad B_0 =- 2^{1/4} {\rm sech}\left({x_0 \over \sqrt{2}}\right)
            = -2^{1/4} ( 1- A_0^2)^{1/2},
\label{match4}
\end{equation}
\begin{equation}
A_1 = {B_0^2 \over 4A_0} + {\rm sech}^2\left({x_0 \over \sqrt{2}}\right)
           {C_1 \over \sqrt{2}}
     = {\sqrt{2}\over 4} {1 - A_0^2 \over A_0} + (1-A_0^2) {C_1\over \sqrt{2}},
\label{match5}
\end{equation}
\begin{equation}
B_1 = {3\over 32} {B_0^3 \over A_0^2} - {\sqrt{2} A_0(1-A_0^2) \over B_0}
                 {C_1\over \sqrt{2}}.
\label{match6}
\end{equation}
Eliminating $C_1$,
\begin{equation}
B_1 = -{\sqrt{2} A_0 A_1 \over B_0} + {3\over 32} {B_0^3\over A_0^2}
      + {1\over 2} {1-A_0^2 \over B_0}.
\label{match7}
\end{equation}
Substituting into our expressions for $H_0(0)$ and $H_1(0)$ from
Eqs.~(\ref{inner10}) and (\ref{inner13a}), we obtain
\begin{equation}
 H_0(0) = 2^{1/4} A_0 ( 1- A_0^2)^{1/2},
\label{match8}
\end{equation}
\begin{equation}
H_1(0) = {2^{3/4} \over 64} {(2A_0^2 +14)(1-A_0^2)^{1/2} \over A_0}
        - {2^{1/4}(2A_0^2 -1) \over (1-A_0^2)^{1/2}} A_1.
\label{match9}
\end{equation}
In order to calculate the superheating field (or, more correctly, the
{\it maximum} superheating field), we need to maximize $H_0(0)$ and
$H_1(0)$ with respect to $A_0$ and $A_1$.
Maximizing $H_0(0)$ with respect to $A_0$, we find that the maximum occurs at
$A_0^* = 1/\sqrt{2}$, $B_0^* = -2^{-1/4}$, so that $H_0 (0) = 2^{-3/4}$.
Substituting this
result into $H_1(0)$, we find the surprising result that the coefficient
of $A_1$ is zero, and $H_1(0) = 2^{3/4} 15/64$.
Our superheating field is then
\begin{equation}
H_{\rm sh} = 2^{-3/4}\kappa^{-1/2} \left[ 1 + {15 \sqrt{2} \over 32} \kappa
            + O(\kappa^2) \right].
\label{final1}
\end{equation}

In order to determine $A_1$ we need to proceed to a higher order
calculation.  The method is the same as before, although the algebra
quickly becomes tedious; we have used the
computer algebra system {\it Maple V} to organize the expansion.
The results from a six term inner expansion are summarized in
Table~\ref{table1}.  Including the next
order term in the expansion in the superheating field, we have
\begin{equation}
H_{\rm sh} = 2^{-3/4}\kappa^{-1/2} \left[ 1 + {15 \sqrt{2} \over 32} \kappa
             - {325\over 1024} \kappa^2 + O(\kappa^3) \right].
\label{final2}
\end{equation}
The first term is exactly the result obtained by the Orsay group
\cite{ginzburg58,orsay}, who used a variational argument to obtain their
result. The second term is identical to the result
obtained by Parr \cite{parr76}.  Parr combined an inspired guess for the
behavior of the order parameter near the surface with a variational
calculation in order to obtain his result.  It is interesting to note
that our result for $A_1$ also agrees with Parr's result.  The advantage
of the method of matched asymptotic expansions is that we can make this
expansion systematic, and therefore in principle carry out this expansion
as far as we wish.  The third term in Eq.~(\ref{final2}) is one of the
new results of this paper; the fourth and fifth terms are included in
Table~\ref{table1}.  With the five-term expansion for $H_{\rm sh}$
it is possible to employ resummation techniques to improve the
expansion.  For instance, the $[2,2]$ Pad\'e approximant \cite{bender} is
\begin{equation}
H_{\rm sh}^{\rm Pade} =  2^{-3/4}\kappa^{-1/2}
  { 1 + 5.4447812\,\kappa + 4.2181012\,\kappa^2 \over
    1 + 4.7818686\, \kappa + 1.3655230\, \kappa^2}.
\label{pade}
\end{equation}
In Fig.~\ref{Hshplot} we compare the numerically calculated superheating
field against the one, two, and three term asymptotic expansions.
The one term (i.e., the Orsay group) result never seems particularly
accurate. There is a marked improvement with the two term expansion, with
the three term expansion offering only a modest additional improvement.
The $[2,2]$ Pad\'e approximant agrees with the numerical data to within about
1\% all the way to $\kappa=1$.

{\it Uniform solutions.}   From the inner and outer expansions it is
possible to construct {\it uniform} solutions, which are asymptotically
correct for all $x$ as $\kappa\rightarrow 0$.  To do this we simply add
the inner and outer solutions of a given order, which guarantees the correct
behavior in the outer region as well as in the boundary layer.  However,
this would produce a result which was $2 f_{\rm match}$  in the
matching region, so we need to subtract $f_{\rm match}$ in order to obtain
the correct behavior in this region.  As an example, we will construct the
2-term uniform solution for the order parameter.  Adding the two-term
outer solution, $f_0(x')+ \kappa f_1(x')$, to the two-term inner solution,
$F_0(X) + \kappa F_1(X)$, subtracting the solution in the matching region,
which is $1/\sqrt{2} + (\sqrt{2}/4)\kappa X - (15/32)\kappa$, and writing
the entire combination in terms of the original variable $x$ (which is the
same as $X$), we obtain
\begin{equation}
f_{\rm unif,2}(x) = \tanh\left( {\kappa x + x_0 \over \sqrt{2}}\right)
- {15\over 16} \kappa\, {\rm sech}^2\left( {\kappa x + x_0 \over
\sqrt{2}}\right)
 + {\kappa \over 4} e^{-\sqrt{2} x}.
\label{unif}
\end{equation}
As $x\rightarrow \infty$, $f_{\rm unif,2}(x)\rightarrow 1$; also,
$f_{\rm unif,2}(0) = 1/\sqrt{2} - (7/32)\kappa$, as we expect.  However,
$f_{\rm unif,2}'(0) = (15/64)\kappa^2$, so that the zero-derivative boundary
condition is only satisfied to $O(\kappa)$.

In Fig.~\ref{matching_fig} and Fig.~\ref{matching_fig2} we compare the
numerically calculated order parameter and magnetic field with the
two term outer solutions and the three term inner solutions.
The agreement is quite good for $\kappa=0.1$, with deviations appearing
at $\kappa=0.5$.  These figures also illustrate the existence of a
matching region where the inner and outer solutions overlap; this
region grows as $\kappa\rightarrow 0$.
Lastly, we show in Fig.~\ref{uniform_fig} how the two term uniform expansion
constructed earlier supplies a uniform approximation to the order parameter and
magnetic field over the whole region.

\section{Stability analysis of the solutions}

Having obtained an asymptotic expansion for the superheating field
$H_{\rm sh}$ in powers of
$\kappa^{1/2}$, we now examine the stability of the solution with respect
to infinitesimal perturbations by studying the second variation
of the free energy, $\delta^2 {\cal F}$.  Perturbations with
$\delta^2 {\cal F}>0$ correspond to stable solutions, while those
with $\delta^2 {\cal F}<0$ correspond to unstable solutions.  We will again use
the method of matched asymptotics to solve for the eigenfunctions of the
linear stability operator.  We first determine the stability in the
simpler one-dimensional situation then we discuss the two-dimensional case.

\subsection{Stability with respect to one-dimensional perturbations}

If we perturb the extremal solution $(f,q)$ of the GL equations by allowing
$f(x)\rightarrow f(x)+\tilde{f}(x)$ and
$q(x)\rightarrow q(x)+\tilde{q}(x)$, then the
second variation of the free energy functional is
\begin{equation}
  \delta^2 {\cal F} =
  \int_0^{\infty}dx\left[\frac{1}{\kappa^2}\tilde{f}'^2+(3f^2+q^2-1)
  \tilde{f}^2+4fq\tilde{f}\tilde{q}+f^2\tilde{q}^2+
  \tilde{q}'^2\right].
\label{stab1}
\end{equation}
The boundary conditions on $\tilde{f}$ and $\tilde{q}$ should be chosen
so as to not perturb $f$ and $h$ at the surface, so that
\begin{equation}
  \tilde{f}'(0) = \tilde{q}'(0)=0, \qquad
          \tilde{f}(\infty)=\tilde{q}(\infty)=0.
\label{stab_boundary}
\end{equation}
We can then integrate Eq.~(\ref{stab1}) by parts to obtain
\begin{equation}
  \delta^2 {\cal F} = \int_0^{\infty}dx\left[\tilde{f}\left(
   -\frac{1}{\kappa^2}\frac{d^2}{dx^2} +
           q^2+3f^2-1\right)\tilde{f} +
  \tilde{q}\left(-\frac{d^2}{dx^2}+
  f^2\right)\tilde{q}+4qf\tilde{q}\tilde{f}\right].
\end{equation}
This quadratic form can be conveniently written as
\begin{equation}
  \delta^2{\cal F} = \int_0^{\infty}dx\, (\tilde{f},\tilde{q})\hat{L}_1
  \left(\begin{array}{c}\tilde{f}\\ \tilde{q}\end{array}\right)
\end{equation}
where $\hat{L}_1$ is the self-adjoint linear operator
\begin{equation}
  \hat{L}_1\left(\begin{array}{c}\tilde{f}\\ \tilde{q}\end{array}\right) =
  \left(\begin{array}{cc}
    -\frac{1}{\kappa^2}\frac{d^2}{dx^2}+q^2+3f^2-1 & 2fq \\
    2fq & -\frac{d^2}{dx^2} +f^2\end{array}\right)
  \left(\begin{array}{c}\tilde{f}\\ \tilde{q}\end{array}\right).
\end{equation}
In order to analyze the stability, expand $\tilde{f}$ and $\tilde{q}$ as
\begin{equation}
\left(\begin{array}{c}\tilde{f}\\ \tilde{q}\end{array}\right) =
\sum_{n} c_n \left(\begin{array}{c}\tilde{f}_n\\ \tilde{q}_n\end{array}\right),
\label{eigen1}
\end{equation}
where the $c_n$'s are real constants, and $(\tilde{f}_n,\tilde{q}_n)$
is a normalized eigenfunction of $\hat{L}_1$ with eigenvalue $E_n$:
\begin{equation}
  \hat{L}_1\left(\begin{array}{c}\tilde{f}_n\\ \tilde{q}_n\end{array}\right) =
  E_n\left(\begin{array}{c}\tilde{f}_n\\ \tilde{q}_n\end{array}\right).
\end{equation}
Then
\begin{equation}
  \delta^2{\cal F} = \sum_n E_n c_n^2.
\end{equation}
The second variation $\delta^2 {\cal F}$ ceases to be positive-definite
when the lowest eigenvalue first becomes negative, indicating that the
corresponding solutions $(f,q)$ of the GL equations are unstable.
Therefore the entire issue of the stability of the solutions has
been reduced to finding the eigenvalue spectrum of the linear stability
operator $\hat{L}_1$, which in the $\kappa\rightarrow 0$ limit can
be studied using matched asymptotic expansions.

{\it Outer solution.} The outer equations for $(\tilde{f},\tilde{q})$
are rescaled with
$x'=\kappa x$ as before to yield (we will drop the subscript $n$
for notational convenience)
\begin{equation}
  -\tilde{f}''+(3f^2+q^2-1)\tilde{f}+2fq\tilde{q} = E\tilde{f},
\label{2d_perturb1}
\end{equation}
\begin{equation}
         -\kappa^2\tilde{q}''+ f^2\tilde{q} + 2fq\tilde{f} = E\tilde{q}.
\label{2d_perturb2}
\end{equation}
Expanding $\tilde{f}$, $\tilde{q}$, and $E$ in powers of $\kappa$, and
recalling that $q=0$ to all orders in $\kappa$ in the outer region,
we have at leading order
\begin{equation}
  -\tilde{f}_0''+(3f_0^2-1)\tilde{f}_0 = E_0\tilde{f}_0,
\label{perturb1}
\end{equation}
where $f_0 = \tanh\left(\frac{x'+x_0}{\sqrt{2}}\right)$.
By changing variables to $y = \tanh\left(\frac{x'+x_0}{\sqrt{2}}\right)$
we see that the solution of Eq.~(\ref{perturb1}) is the associated
Legendre function of the first kind:
\begin{equation}
  \tilde{f}_0(x') = c_0 P_2^{\mu}\left[\tanh\left(\frac{x'+x_0}{\sqrt{2}}
             \right)\right],
\label{legendre}
\end{equation}
where $\mu = -\sqrt{2(2-E_0)}$ and $c_0$ is a constant.
The leading order solution for $\tilde{q}$ is $\tilde{q}_0=0$.

{\it Inner solution.} To obtain the inner equations, we rescale as in
Eq.~(\ref{inner1}), with the perturbations rescaled as
\begin{equation}
\tilde{f}(x') = \tilde{F}(X),\quad \tilde{q}(x') = \kappa^{-1/2}\tilde{Q}(X),
\label{perturb2}
\end{equation}
such that
\begin{eqnarray}
  -\frac{1}{\kappa^2}\tilde{F}''+
  (3F^2+\frac{1}{\kappa}Q^2-1)\tilde{F}+\frac{1}{\kappa}2 FQ
\tilde{Q} = E\tilde{F}, \\
         -\tilde{Q}''+ F^2\tilde{Q} + 2FQ\tilde{F} = E\tilde{Q}.
  \end{eqnarray}
To leading order, $\tilde{F_0}'' = 0$, so that $\tilde{F_0} = a_0$, with
$a_0$ a constant.  The leading order equation for
the variation in $Q$ is
\begin{equation}
  -\tilde{Q}_0'' + 2F_0Q_0\tilde{F}_0+(F_0^2-E_0)\tilde{Q}_0 = 0.
\end{equation}
The solution which satisfies the boundary condition $\tilde{Q}'(0)=0$ is
\begin{equation}
  \tilde{Q}_0(X) =
  \frac{2a_0A_0B_0}{E_0}\left(e^{-A_0X}-\frac{A_0}{\sqrt{A_0^2-E_0}}
e^{-\sqrt{A_0^2-E_0}X}\right).
\end{equation}
At $O(\kappa)$ we find
\begin{equation}
  \tilde{F}_1'' = Q_0^2\tilde{F}_0+2F_0Q_0\tilde{Q}_0,
\end{equation}
with the solution
\begin{equation}
  \tilde{F}_1(X) =\displaystyle
  a_1+ a_0B_0^2\left[\frac{E_0+4A_0^2}{4A_0^2E_0}e^{-2A_0X}-
    \frac{4A_0^2}{E_0}\frac{e^{-(A_0+\sqrt{A_0^2-E_0})X}}
            {(A_0+\sqrt{A_0^2-E_0})^2\sqrt{A_0^2-E_0}}\right]
   \qquad\atop\quad\displaystyle
     +a_0B_0^2\left[\frac{E_0+4A_0^2}{2A_0E_0}-
            \frac{4A_0^3/E_0}{(A_0+\sqrt{A_0^2-E_0})\sqrt{A_0^2-E_0}}\right]X.
\end{equation}
We now have enough terms in the inner and outer region for a nontrivial match.

{\it Matching.} We complete the matching of the inner and outer perturbations
to obtain the eigenvalue, $E_0$.  Performing a two term inner expansion of
the one term outer solution, we have
\begin{equation}
  \tilde{f}_0(\kappa X) \sim c_0 \left[ P_2^{\mu}(A_0) +
  \frac{1}{\sqrt{2}}\mbox{sech}^2(x_0/\sqrt{2})
   \frac{dP_2^{\mu}(A_0)}{d A_0 } \kappa X \right],
\end{equation}
where we have used $\tanh(x_0/\sqrt{2}) = A_0$.
Next, the one term outer expansion of the two term inner solution is
\begin{equation}
  \tilde{F}_0(x'/\kappa) + \kappa \tilde{F}_1(x'/\kappa) \sim
 a_0+\frac{a_0 2^{1/2} (1-A_0^2)}{E_0}\left[
\frac{E_0+4A_0^2}{2A_0}
      -\frac{4A_0^3}{(A_0+\sqrt{A_0^2-E_0})\sqrt{A_0^2-E_0}}\right] x',
\end{equation}
where we have used $B_0 = -2^{1/4}(1-A_0^2)^{1/2}$.
Matching the two expansions using the van Dyke matching principle yields
\begin{equation}
  c_0 = \frac{a_0}{P_2^{\mu}(A_0)},
\end{equation}
\begin{equation}
  \frac{1}{P_2^{\mu}(A_0)}\frac{dP_2^{\mu}(A_0)}{dA_0} =
  \frac{2}{E_0}\left[
    \frac{E_0+4A_0^2}{2A_0}
  -\frac{4A_0^3}{(A_0+\sqrt{A_0^2-E_0})\sqrt{A_0^2-E_0}}\right].
\label{metastable}
\end{equation}
The last equation is a rather complicated implicit equation for the
eigenvalue $E_0(A_0)$, which generally must be solved numerically.
However, when $A_0=1/\sqrt{2}$ we find $E_0=0$, corresponding to the
critical case, with $E>0$ for $A_0>1/\sqrt{2}$.  The numerical evaluation
of Eq.~(\ref{metastable}) is shown in Fig.~\ref{stable_plot}.  Therefore, we
see that our maximum superheating field (at lowest order) corresponds to
the limit of metastability for these one-dimensional perturbations.
In Fig.~\ref{nose} we show $A_0$ as a function of the lowest order
magnetic field at the surface, $H_0$, from Eq.~(\ref{match8}).
The stability analysis of this section shows that only the upper
branch of this double valued function corresponds to solutions which
are locally stable, with the field at the ``nose'' being the superheating
field.

\subsection{Stability with respect to two-dimensional perturbations}

We next turn to the stability of the solutions with respect to two
dimensional perturbations. If we perturb the extremal solution $(f,{\bf q})$
of the GL equations by allowing
$f\rightarrow f+\delta f$ and ${\bf q}\rightarrow {\bf q}+\delta {\bf q}$,
then the second variation of the free energy functional is
\begin{equation}
  \delta^2 {\cal F} = \int dx\,dy \left[\frac{1}{\kappa^2}
({\bf \nabla}\delta f)^2+
  4 f (\delta f) {\bf q}\cdot\delta {\bf q}+f^2(\delta {\bf q})^2+
  (3 f^2+{\bf q}^2-1)(\delta f)^2 + ({\bf\nabla}\times\delta {\bf q})^2\right]
\label{2d}
\end{equation}
(we neglect perturbations along the $z$-direction).
Expanding in Fourier modes with respect to $y$ \cite{kramer68},
\begin{equation}
  \delta f(x,y) = \tilde{f}(x)\cos ky, \quad
  \delta q_x(x,y) = \tilde{q}_x(x)\sin ky,\quad
  \delta q_y (x,y) = \tilde{q}_y(x)\cos ky,
\end{equation}
substituting into Eq.~(\ref{2d}), recalling that ${\bf q} = (0,q(x),0)$,
and integrating over $y$, we obtain (up to a multiplicative constant)
\begin{equation}
  \delta^2{\cal F} =
  \int_0^{\infty}dx\left[\frac{1}{\kappa^2}\tilde{f}'^2+(3f^2+q^2
    +\frac{1}{\kappa^2}k^2-1)
  \tilde{f}^2+4fq\tilde{f}\tilde{q}_y+f^2(\tilde{q}_x^2+\tilde{q}_y^2)+
  (\tilde{q}_y'-k\tilde{q}_x)^2\right].
\end{equation}
By integrating by parts and using the boundary conditions,
Eq.~(\ref{stab_boundary}), we can cast this functional into the form
\begin{equation}
  \delta^2{\cal F} = \int_0^{\infty}dx\, (\tilde{f},\tilde{q}_y,\tilde{q}_x)
\hat{L}_2 \left(\begin{array}{c}\tilde{f}\\ \tilde{q}_y\\
\tilde{q}_x\end{array}
  \right),
\end{equation}
where the self-adjoint linear operator $\hat{L}_2$ is given by
\begin{equation}
  \hat{L}_2\left(\begin{array}{c}\tilde{f}\\ \tilde{q}_y\\ \tilde{q}_x
 \end{array}\right) =
  \left(\begin{array}{ccc}
    -\frac{1}{\kappa^2}\frac{d^2}{dx^2}+q^2+3f^2+ k^2/\kappa^2-1 & 2fq & 0\\
    2fq & -\frac{d^2}{dx^2} +f^2 & - k \frac{d}{dx}\\
   0 &  k \frac{d}{dx} & f^2 + k^2 \end{array}\right)
\left(\begin{array}{c}\tilde{f}\\ \tilde{q}_y \\ \tilde{q}_x
\end{array}\right).
\end{equation}
As in the previous section, we want to determine the eigenvalue spectrum of
this operator. We are primarily interested in the effects of long-wavelength
perturbations (i.e., $k\rightarrow 0$), so we rescale $k$ as $k=\kappa k'$.
Then the eigenvalue
equations in terms of the outer coordinate $x'=\kappa x$ are (dropping the
prime on $k$ from now on)
\begin{equation}
  -\tilde{f}''+(3f^2+q^2-1+k^2)\tilde{f}+2fq\tilde{q} = E\tilde{f},
\label{2d1}
\end{equation}
\begin{equation}
 - \kappa^2 \tilde{q}_y'' + f^2 \tilde{q}_y + 2 fq\tilde{f}
            - \kappa^2 k \tilde{q}_x' = E\tilde{q}_y,
\label{2d2}
\end{equation}
\begin{equation}
 \kappa^2 k \tilde{q}_y' + (f^2 + \kappa^2 k^2) \tilde{q}_x = E \tilde{q}_x.
\label{2d3}
\end{equation}
By using the last equation we may eliminate $\tilde{q}_x$ from Eq.~(\ref{2d2}),
which becomes
\begin{equation}
- \kappa^2 {d \over dx}\left[ {f^2 - E \over f^2 + \kappa^2 k^2 - E}
 \tilde{q}_y'\right] + f^2\tilde{q}_y + 2fq\tilde{f} = E \tilde{q}_y.
\label{2d4}
\end{equation}
For $k=0$ Eqs.~(\ref{2d1}) and (\ref{2d4}) reduce to the one-dimensional
perturbation equations of the last section, Eqs.~(\ref{2d_perturb1})
and (\ref{2d_perturb2});  for $E=0$ they reduce to
the Euler-Lagrange equations derived by Kramer \cite{kramer68}.

The perturbation equations (\ref{2d1}) and (\ref{2d4}) may be solved
by the method of matched asymptotic expansions, just as in the
one-dimensional case.  The derivation of the eigenvalue condition
is essentially identical, with the final result that
\begin{equation}
  \frac{1}{P_2^{\mu}(A_0)}\frac{dP_2^{\mu}(A_0)}{dA_0} =
  \frac{2}{E_0}\left[
    \frac{E_0+4A_0^2}{2A_0}
  -\frac{4A_0^3}{(A_0+\sqrt{A_0^2-E_0})\sqrt{A_0^2-E_0}}\right],
\end{equation}
where now $\mu = - \sqrt{2(2+E_0-k^2)}$. The eigenvalue $E_0(k)$ is plotted
in Fig. \ref{stable_plot2} for several different values of $A_0$.
For $A_0>1/\sqrt{2}$, $E_0(k)>0$ for all $k$, while for $A_0<1/\sqrt{2}$
there exists a band of long-wavelength perturbations for which
$E_0(k)<0$.  In all cases the most unstable modes are at $k=0$, i.e.,
the one-dimensional perturbations are the least stable.  This is
in contrast to the large-$\kappa$ limit, where the most unstable mode
occurs for $k\neq 0$ \cite{galaiko66,kramer68,chapman95}.

\section{Nonlocal effects as $\kappa \rightarrow 0$}

In the previous sections we have studied superheating in type-I
superconductors starting from the conventional GL
equations, which assume a {\it local} relationship between the current
and the vector potential.  However, in very clean type-I superconductors
{\it nonlocal} effects are often important (in the {\it Pippard} limit;
see Ref.~\cite{tinkham}).  We can model these effects by replacing the
second GL equation, Eq.~(\ref{GL5}), by a nonlocal equation
of the form
\begin{equation}
\kappa^2 q'' - \int_{0}^{\infty} K(x - x') f^2(x') q(x')\, dx' = 0,
\label{non1}
\end{equation}
where $K(x-x')$ is a kernel whose Fourier transform $K(k)$ behaves as
\begin{eqnarray}
K(k) = \left\{
  \begin{array}{ll}
   \lambda^2/\lambda_{L}^2\qquad {\rm (local\ limit)}; \\
   a/|k| \qquad {\rm (extreme\ anomalous\ limit)}, \\
   \end{array}
   \right.
\label{limits}
\end{eqnarray}
with $\lambda_L$ the London penetration depth and $a$ a constant
\cite{tinkham}.   For $\lambda\approx \lambda_L$ we recover the local limit
considered in the previous sections of this paper. It is still possible
to calculate the
superheating field in this nonlocal limit using the method of matched
asymptotic expansions.  Indeed, the prescription is the same as for the
local case discussed above; we only need to solve a slightly more
complicated inner problem.  In this section we will calculate the
leading order superheating field in the nonlocal limit, in order to
further illustrate the power and flexibility of our method.

{\it Outer solution.} The outer solution is the same as before; the
vector potential is zero to all orders, and the first two terms in the
expansion for the order parameter are given by Eqs.~(\ref{outer3}) and
(\ref{outer7}).

{\it Inner solution.}  In the inner region we rescale the variables as
in Eq.~(\ref{inner1}).  In terms of these variables Eq.~(\ref{non1})
becomes
\begin{equation}
Q'' - \int_{0}^{\infty} K(X - X') F^2(X') Q(X')\, dX' = 0.
\label{non2}
\end{equation}
We need to solve this equation, along with the first GL
equation, Eq.~(\ref{inner2}), perturbatively in $\kappa$.  Expanding
$F$ and $Q$ as in Eqs.~(\ref{inner5})--(\ref{inner7}), we obtain
$F_0(X)=A_0$, as before, and
\begin{equation}
Q_0'' - A_0^2 \int_{0}^{\infty} K(X-X') Q_0 (X')\, dX' = 0.
\label{non3}
\end{equation}
This is an integral equation of the Wiener-Hopf type \cite{integral}.
To solve, we Fourier transform, introducing
\begin{equation}
Q_{+}(k) = \int_{0}^{\infty} Q_0(X) e^{ikX}\, dX.
\label{wh1}
\end{equation}
After Fourier transforming the integral equation, we perform a
Wiener-Hopf factorization \cite{integral}, with the result that
\begin{equation}
Q_{+}(k) = B_0 { i e^{i\varphi(k)} \over [ k^2 + A_0^2 K(k)]^{1/2}},
\label{wh2}
\end{equation}
where $B_0 = Q_0(0)$ is a constant, and
\begin{equation}
\varphi(k) = {k\over \pi} \int_{0}^{\infty} \ln\left[ { k^2 + A_0^2 K(x)
           \over x^2 + A_0^2 K(k) }\right] {1\over x^2 - k^2} \, dx.
\label{wh3}
\end{equation}
The Fourier transform can be inverted once a particular form for $K(k)$ is
specified (although this is unnecessary for the calculation of the
superheating field; see below).  The magnetic field at this order is
$H_0(0)=-A_0 B_0$ as before.

Proceeding to the next order, we have
\begin{equation}
F_1'' = A_0 Q_0^2.
\label{non4}
\end{equation}
By applying the boundary condition $F_1'(0)=0$, we find the general solution
\begin{equation}
F_1(X) = A_1 + A_0\left[ X\int_{0}^{X} Q_0^2(y)\, dy
              - \int_{0}^{X} y Q_0^2(y)\, dy\right],
\label{non5}
\end{equation}
with $A_1 = F_1(0)$ another constant.
The equation for $Q_1(X)$ is a rather messy inhomogeneous Wiener-Hopf
integral equation.  Fortunately, its solution is not needed for the
leading order calculation of the superheating field.

{\it Matching.} We now turn to the matching of the inner and outer solutions.
The two term inner expansion of the one term outer solution is
\begin{equation}
f_0(\kappa X) \sim \tanh\left( {x_0 \over \sqrt{2}}\right)
    + {\kappa \over \sqrt{2}} {\rm sech}^2\left( {x_0 \over \sqrt{2}}\right)X.
\label{non6}
\end{equation}
The one term outer expansion of the two term inner solution is
\begin{equation}
F_0(x'/\kappa) + \kappa F_1(x'/\kappa) \sim  A_0 + A_0 \left( \int_{0}^{\infty}
        Q_0^2(y) \, dy \right) x' .
\label{non7}
\end{equation}
By using the van Dyke matching principle we find
$A_0 = \tanh(x_0/\sqrt{2})$ as before, and
\begin{equation}
A_0 \int_0^{\infty} Q_0^2(y)\, dy = {1\over \sqrt{2}} (1-A_0^2).
\label{non8}
\end{equation}
We can use Parseval's identity to express the left hand side of
Eq.~(\ref{non8}) in terms of $|Q_{+}(k)|^2$, and then use
Eq.~(\ref{wh2}) to finally arrive at
\begin{equation}
{A_0 B_0^2\over 2\pi}  \int_{-\infty}^{\infty} {1\over k^2 + A_0^2 K(k)}\, dk
  = {1\over \sqrt{2}} (1-A_0^2).
\label{non9}
\end{equation}

To calculate the superheating field we use Eq.~(\ref{non9}) to express
$B_0$ as a function of $A_0$; we then substitute
this result into $H_0(0) = -A_0 B_0$ and maximize with respect to
$A_0$ in order to determine the lowest order superheating field.
In the local limit, $K(k) = \lambda^2/\lambda_L^2$,
and we obtain $A_0^* = 1/\sqrt{2}$ and
$H_0(0)= 2^{-3/4} (\lambda_L/\lambda)$, which is the same as our previous
result when $\lambda\approx \lambda_L$.  In the extreme anomalous limit
$K(k) = a/|k|$, with $a=(3\pi/4) (\lambda^2 \xi /\lambda^2_L \xi_0)$
in the Pippard theory \cite{tinkham}, where $\xi_0$ is the zero temperature
coherence length.  Performing the integral, we find
\begin{equation}
B_0 = - 3^{3/4} 2^{-3/4} a^{1/6} A_0^{-1/6} (1-A_0^2)^{1/2},
\label{non10}
\end{equation}
so that
\begin{equation}
H_0(0) = 3^{3/4} 2^{-3/4} a^{1/6} A_0^{5/6} (1-A_0^2)^{1/2}.
\label{non11}
\end{equation}
The maximum occurs at $A_0^* = \sqrt{5/11}$, so that
$H_0(0) = 0.721\, a^{1/6}$. Therefore, the superheating field is
\begin{equation}
H_{\rm sh} = 0.721\, a^{1/6} \kappa^{-1/2} + O(\kappa^{1/2})
\qquad {(\rm extreme\ anomalous\ limit)}.
\label{non12}
\end{equation}
The same result has been obtained by Smith {\it et al.} \cite {smith70}
using an approximation for the order parameter along with a variational
calculation (in the spirit of method used by the Orsay group \cite{orsay}).
The advantage of our method is that it can be systematically improved upon.
Although we have not checked the stability in the extreme anomalous limit,
the procedure should be entirely analogous to that of the previous section.

\section{Large-$\kappa$ results}

So far we have used the method of matched asymptotics to solve the
GL equations in the small-$\kappa$ limit.  Chapman \cite{chapman95}
has recently used the same method to treat the one dimensional
GL equations in the high-$\kappa$ limit.
His final result for the superheating field is
\begin{equation}
H_{\rm sh} = \frac{1}{\sqrt{2}}+C\kappa^{-4/3} + O(\kappa^{-6/3})
\label{chapsh1}
\end{equation}
where the constant $C$ is determined from the solution of the second
Painlev\`e transcendent; a numerical evaluation yields $C=0.326$
\cite{dolgert}.  The first term was originally derived by Ginzburg
\cite{ginzburg58}, and the second term with the unusual dependence upon
$\kappa$ is the new term.  As seen in Fig~\ref{chapman} the asymptotic
and numerical results agree very well.   It turns out, however, that the
calculated $H_{\rm sh}$ is not actually the
superheating field, since the one dimensional solution in the large-$\kappa$
limit is unstable with respect to two-dimensional perturbations
\cite{galaiko66,kramer68,chapman95}; these instabilities occur at
at the smaller field $H_{\rm sh}^{2D} = \sqrt{5}/3\sqrt{2}=0.527$.
This situation is quite different from that of the small- $\kappa$ limit
in which our stability calculation (Section IV) found the limit of
stability to be right at $H_{\rm sh}$.

\section{Discussion}

In this paper, the one-dimensional GL equations are solved analytically
and numerically for a semi-infinite superconducting sample in the
small-$\kappa$ limit in order to determine the maximum superheating field
$H_{\rm sh}$.  We have used the method of matched asymptotic expansions
to construct for the first time a systematic perturbative solution of
the Ginzburg-Landau equations, the results of which agree quite closely with
our numerical solutions.  The same method has been used to
determine the stability of these solutions with respect to both one-
and two-dimensional infinitesimal fluctuations; our analysis shows
that two dimensional fluctuations do not lead to any additional
destabilizing effects, in contrast to the situation in the large-$\kappa$
limit. With little modification this method can also be adapted to treat
nonlocal electrodynamic effects.  Finally, our numerical results for
large-$\kappa$  compare well with Chapman's asymptotic analysis of this
regime.  Taken collectively, our results demonstrate the effectiveness
of the method of matched asymptotic expansions for dealing with
boundary layer problems in the theory of superconductivity.  We
hope that others will find useful applications of the methods
developed in this paper in treating inhomogeneous superconductors.

\acknowledgments
We would like to thank Dr. Chung-Yu Mou for helpful discussions,
Dr. S. J. Chapman for useful correspondence,  and Dr. C. Bolley for
sending us her preprints and reprints on superheating
in superconductors. This work was supported by   NSF Grant DMR 92-23586,
as well as by the Alfred P. Sloan Foundation (ATD).

\begin{figure}
\caption{A comparison of the numerically calculated superheating field
$H_{\rm sh}$ (heavy line) with the three term asymptotic expansion for
small-$\kappa$, and the $[2,2]$ Pad\'e approximant.  The one-term
expansion due to the Orsay group deviates
systematically from the calculated superheating field.  The two- and
three-term expansions provide a marked improvement over the one-term
expansion.}
\label{Hshplot}
\end{figure}

\begin{figure}
\caption{A comparison of the three term inner and outer solutions for the
order parameter and the magnetic field
with the numerical solution for $\kappa=0.1$.  The asymptotic solutions
approximate the computed values only in the appropriate regions.
The matching region where the inner and outer meet is
$O(\kappa)$ as can be estimated from the inner solution for $f$.}
\label{matching_fig}
\end{figure}

\begin{figure}
\caption{The same as Fig. \ref{matching_fig} for $\kappa=0.5$.}
\label{matching_fig2}
\end{figure}

\begin{figure}
\caption{A comparison of the two-term uniform solution for the
order parameter, $f_{\rm unif,2}(x)$ (dashed line),
with the numerical solution (solid line) at
$\kappa= 0.5$.  The disagreement of the uniform solution with the boundary
condition at $x=0$ is of order $\kappa^2$.}
\label{uniform_fig}
\end{figure}

\begin{figure}
\caption{The stability eigenvalue $E(A_0)$, with $A_0$ the value of the
order parameter at the surface at leading order.  We see that $E>0$
for $A_0> 1/\protect\sqrt{2}$, indicating locally stable solutions.}
\label{stable_plot}
\end{figure}

\begin{figure}
\caption{The order parameter at the surface, $A_0$, as a function of the
field at the surface, $H_0$, at leading order.  The stability analysis
shows that only the upper branch corresponds to locally stable solutions.
The field at the ``nose'' is the limit of stability, and corresponds
to the superheating field $H_0 = 2^{-3/4}=0.595$.}
\label{nose}
\end{figure}

\begin{figure}
\caption{The stability eigenvalue $E(k)$ for two-dimensional perturbations
of wavenumber $k$, for several different values of $A_0$. For
$A_0> 1/\protect\sqrt{2}$ the eigenvalue is stable for all wavenumbers,
while for $A_0< 1/\protect\sqrt{2}$ there exists a band of wavenumbers
for which the solution is unstable. }
\label{stable_plot2}
\end{figure}

\begin{figure}
\caption{The numerically calculated superheating field for large-$\kappa$
(solid line), compared with the two-term asymptotic expansion derived by
Chapman (dashed line).  The slope of the dashed line is $-4/3$.}
\label{chapman}
\end{figure}

\begin{table}
\caption{Summary of the results of the small-$\kappa$ expansion for the
superheating field.  Here $A_n$ is the value of the order parameter $F(X)$
at the surface at $n$-th order, $B_n$ is the value of the vector potential
$Q(X)$ at $n$-th order, $C_n$ is the coefficient of the $n$-th term in the
outer expansion of the order parameter, and $H_n(0)$ is the $n$-th order
term in the expansion of the superheating field.}

\label{table1}

\begin{tabular}{ccccc}

$n$     & $A_n$     & $B_n$     & $C_n$      & $H_n(0)$  \\ \tableline \\

0 & $2^{-1/2}$ & $-2^{-1/4}$      &  $1$   &  $2^{-3/4}$ \\
1 & $-7/32   $ & $-(9/16)2^{1/4}$ &  $-(15/16)2^{1/2}$ & $(15/64)2^{3/4}$ \\
2 & $-(17/1024)2^{1/2}$& $(159/2048)2^{3/4}$& $225/256$ &
$-(325/2048)2^{1/4}$\\
3 & $3211/16384$ & $-(745/4096)2^{1/4}$ &$-(1125/4096)2^{1/2}$&
                                    $ (14191/65536) 2^{3/4}$ \\
4& $-(623575/1572864)2^{1/2}$ & $(16223049/20971520)2^{3/4}$ &
                 $16875/131072$ & $-(78495727/62914560)2^{1/4}$ \\

\end{tabular}
\end{table}

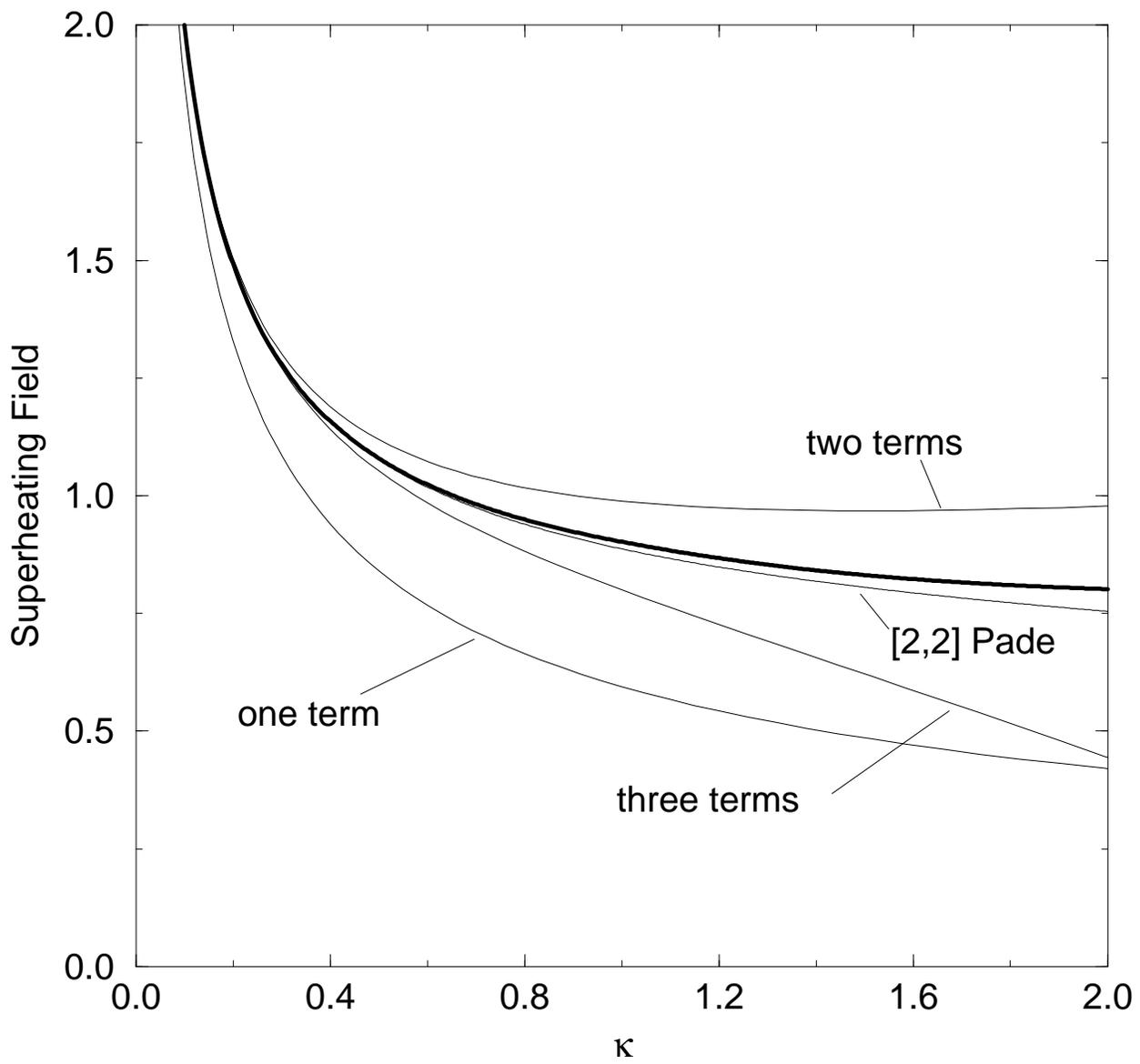
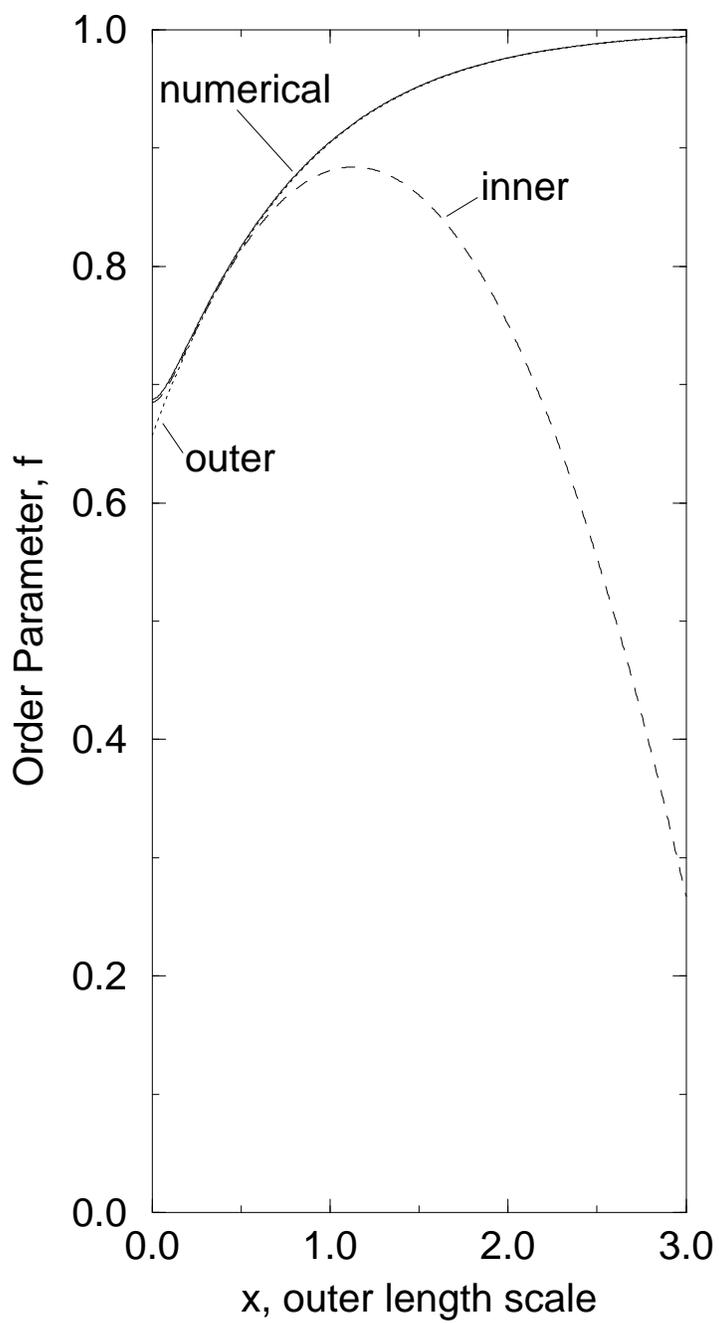
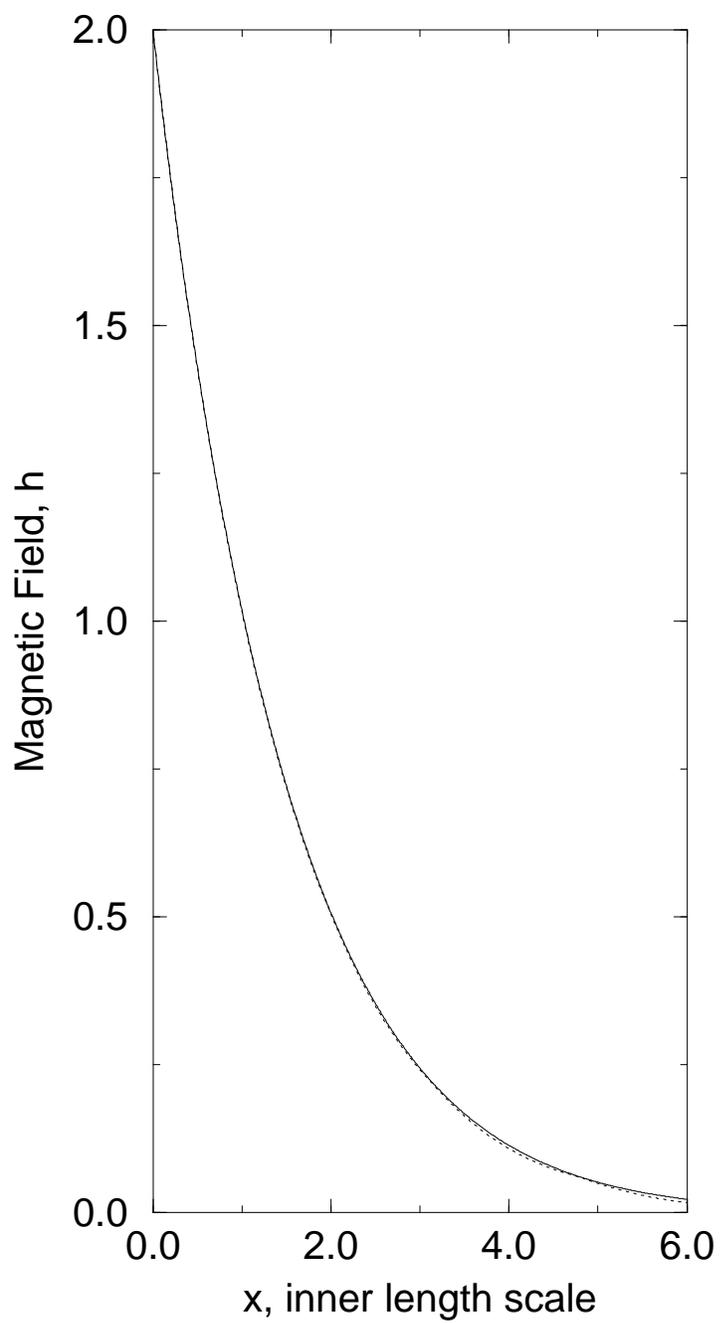
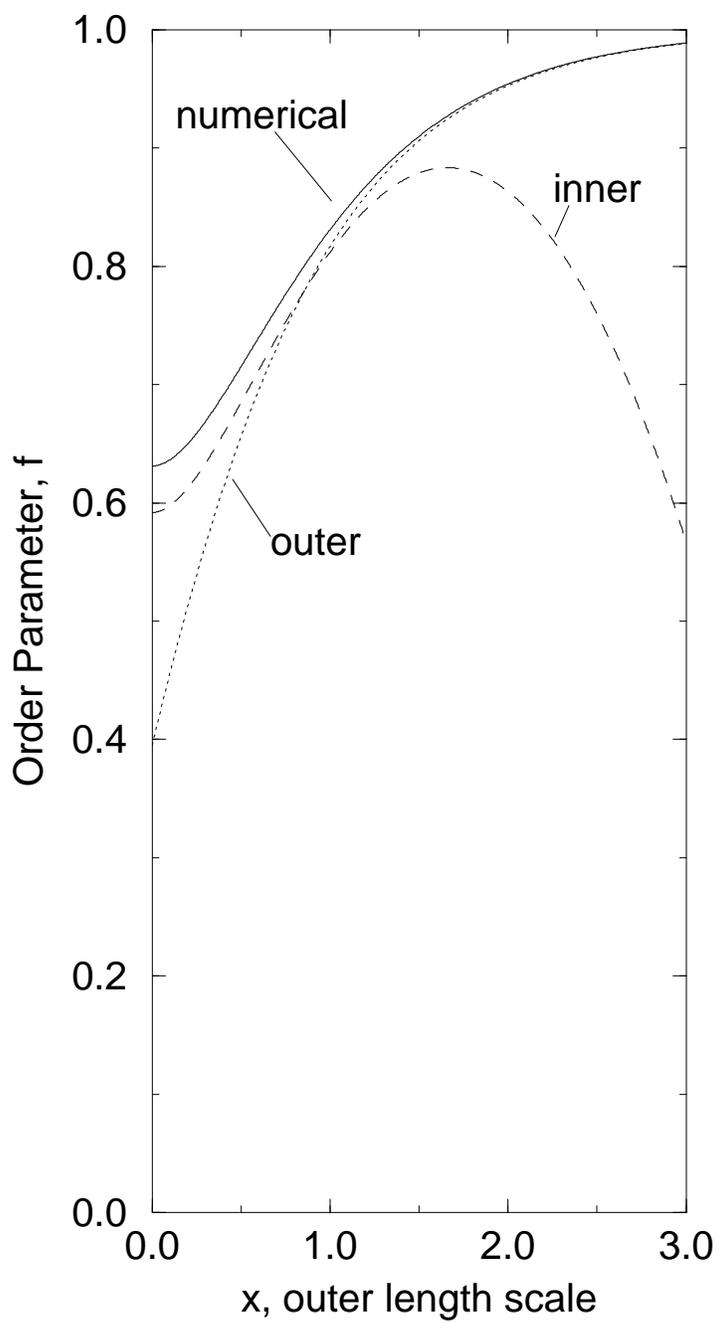
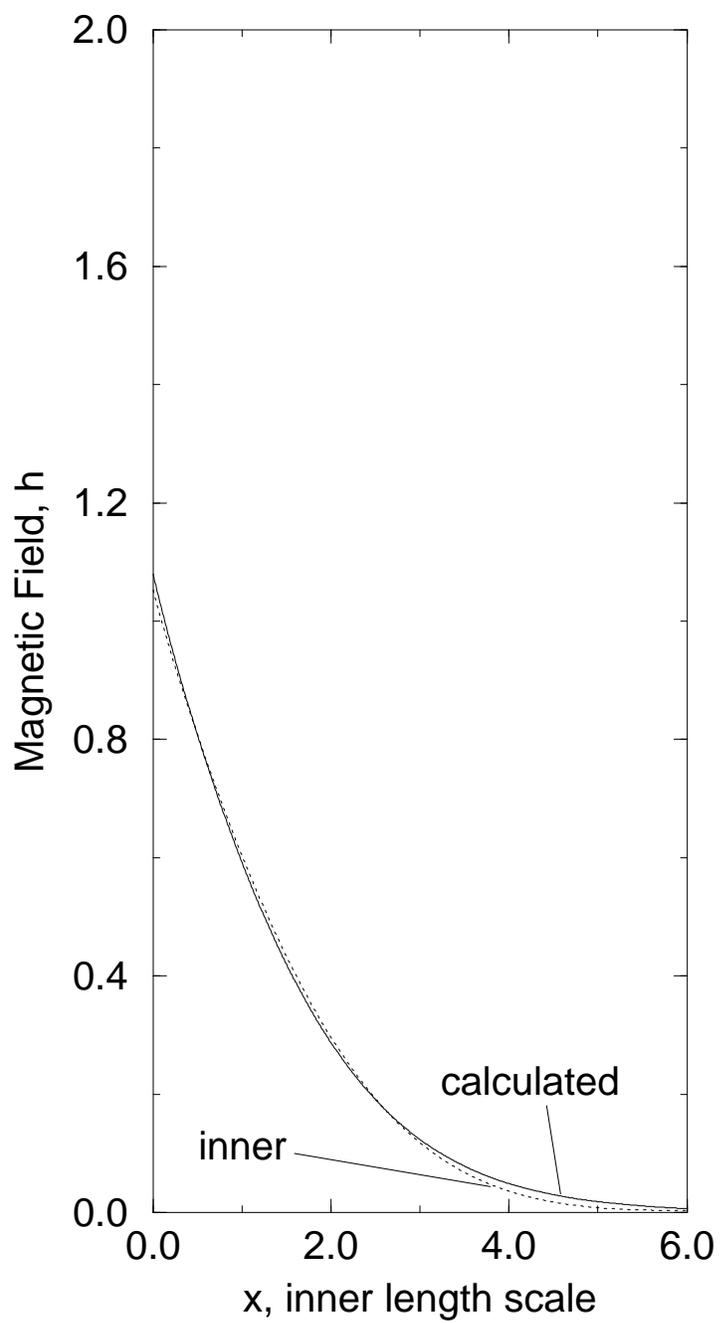
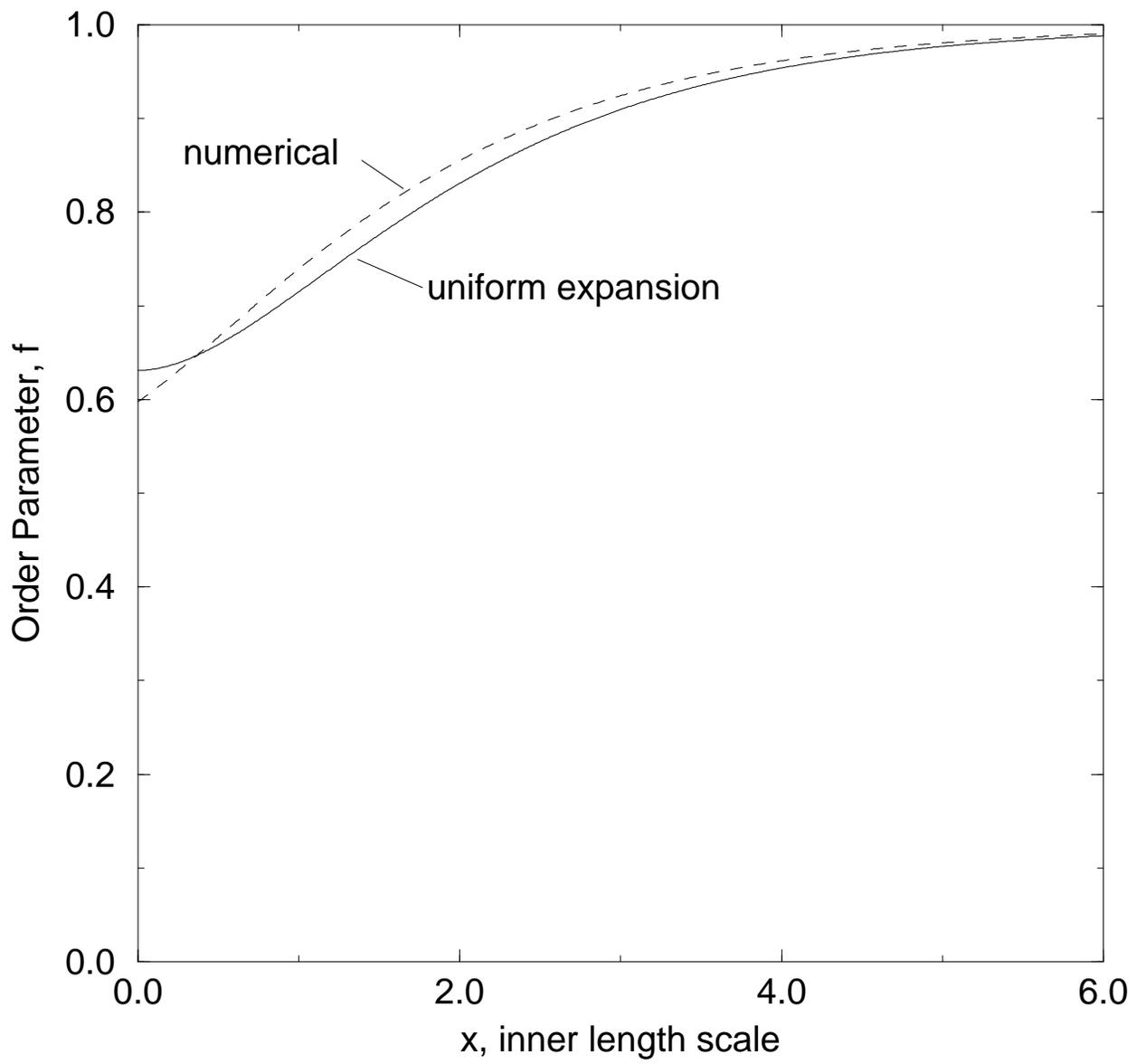
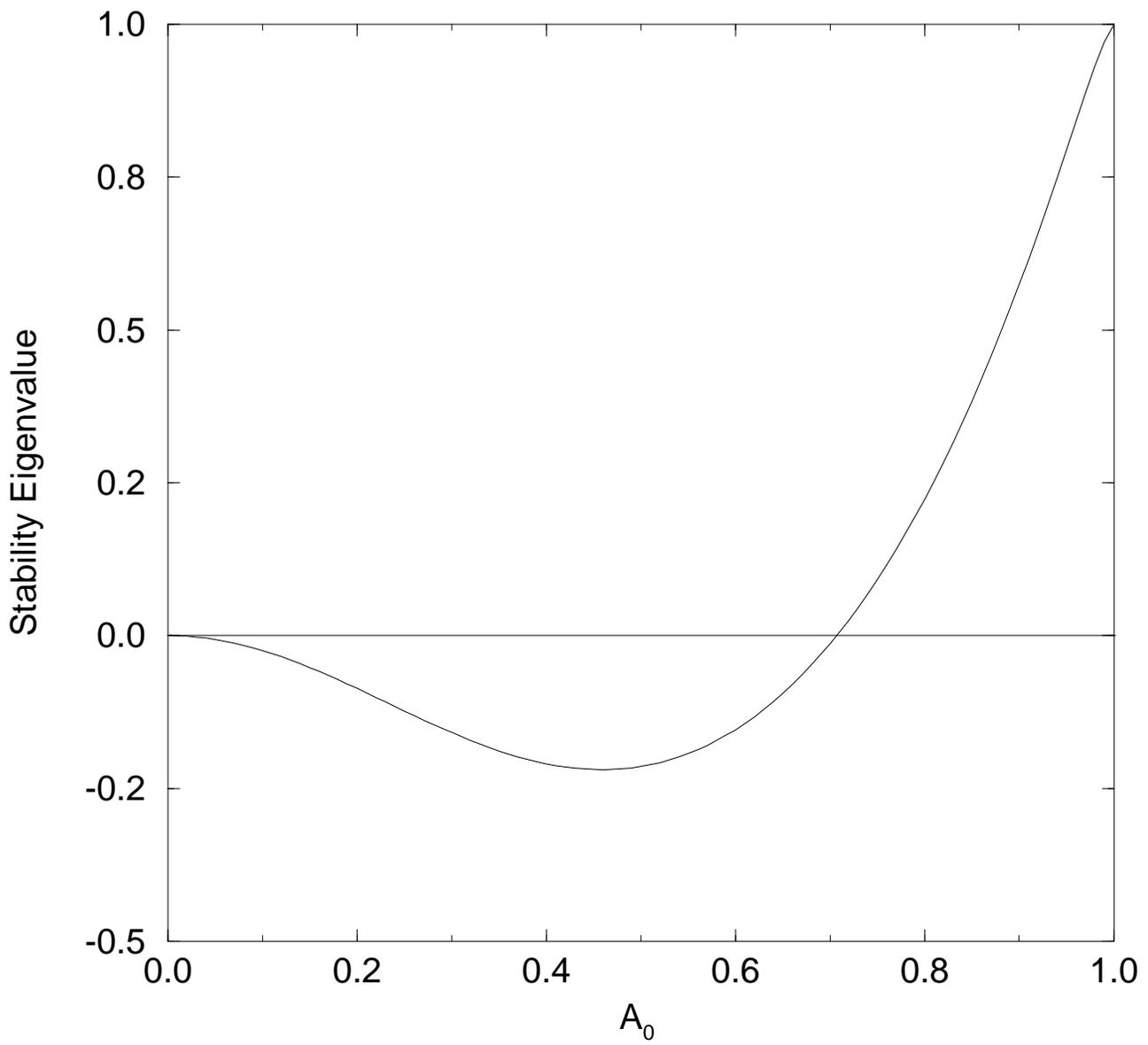
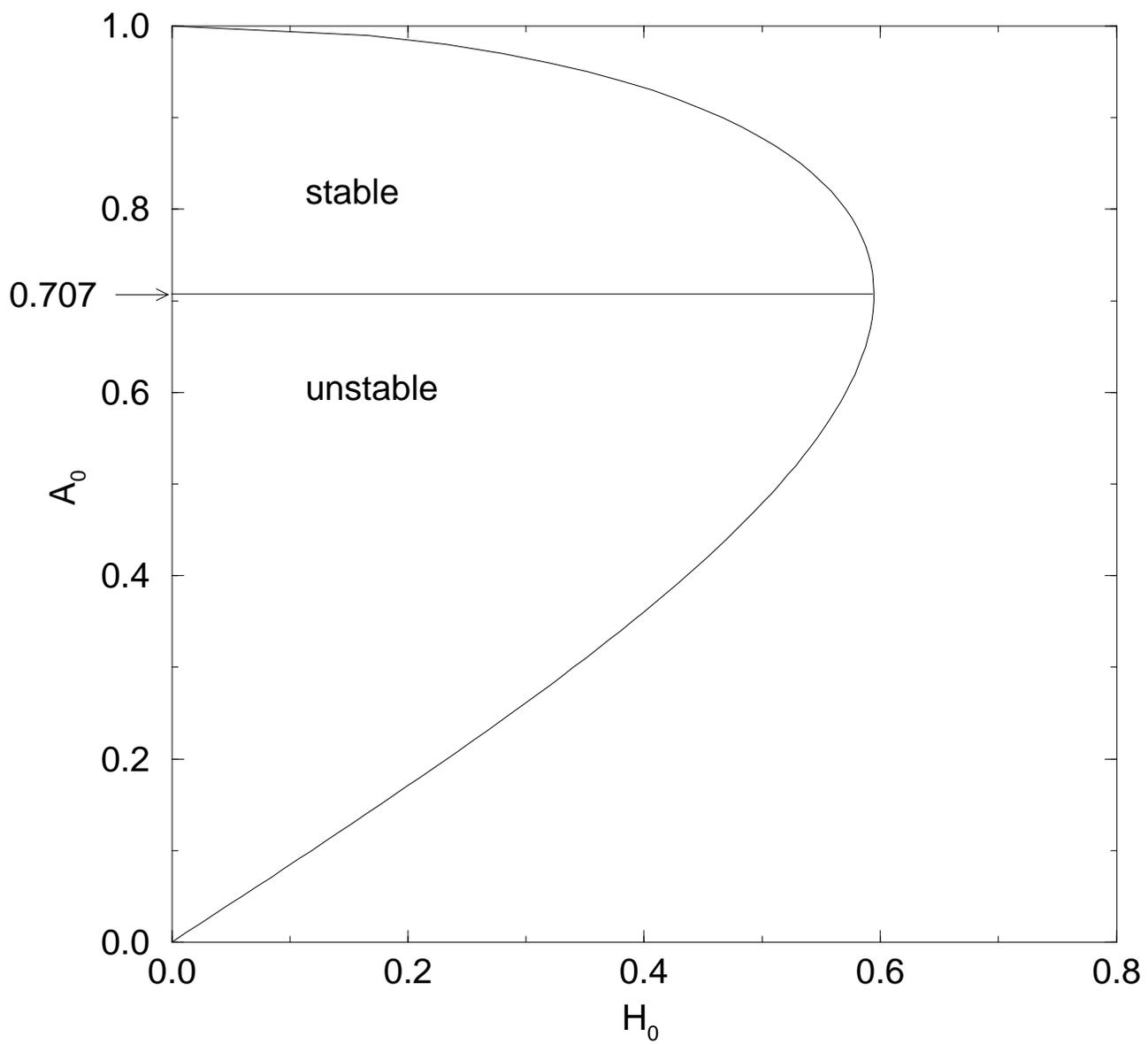
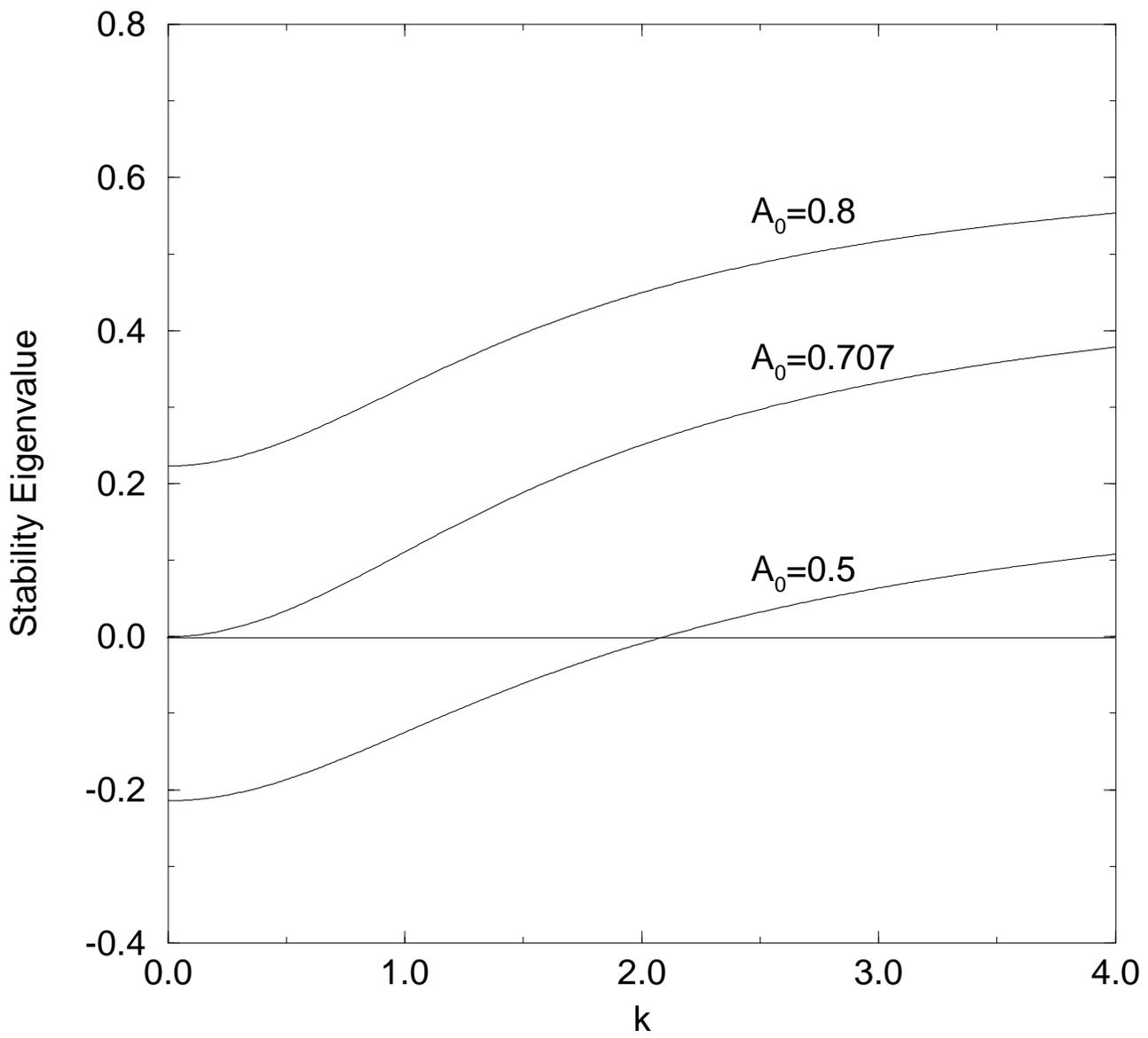
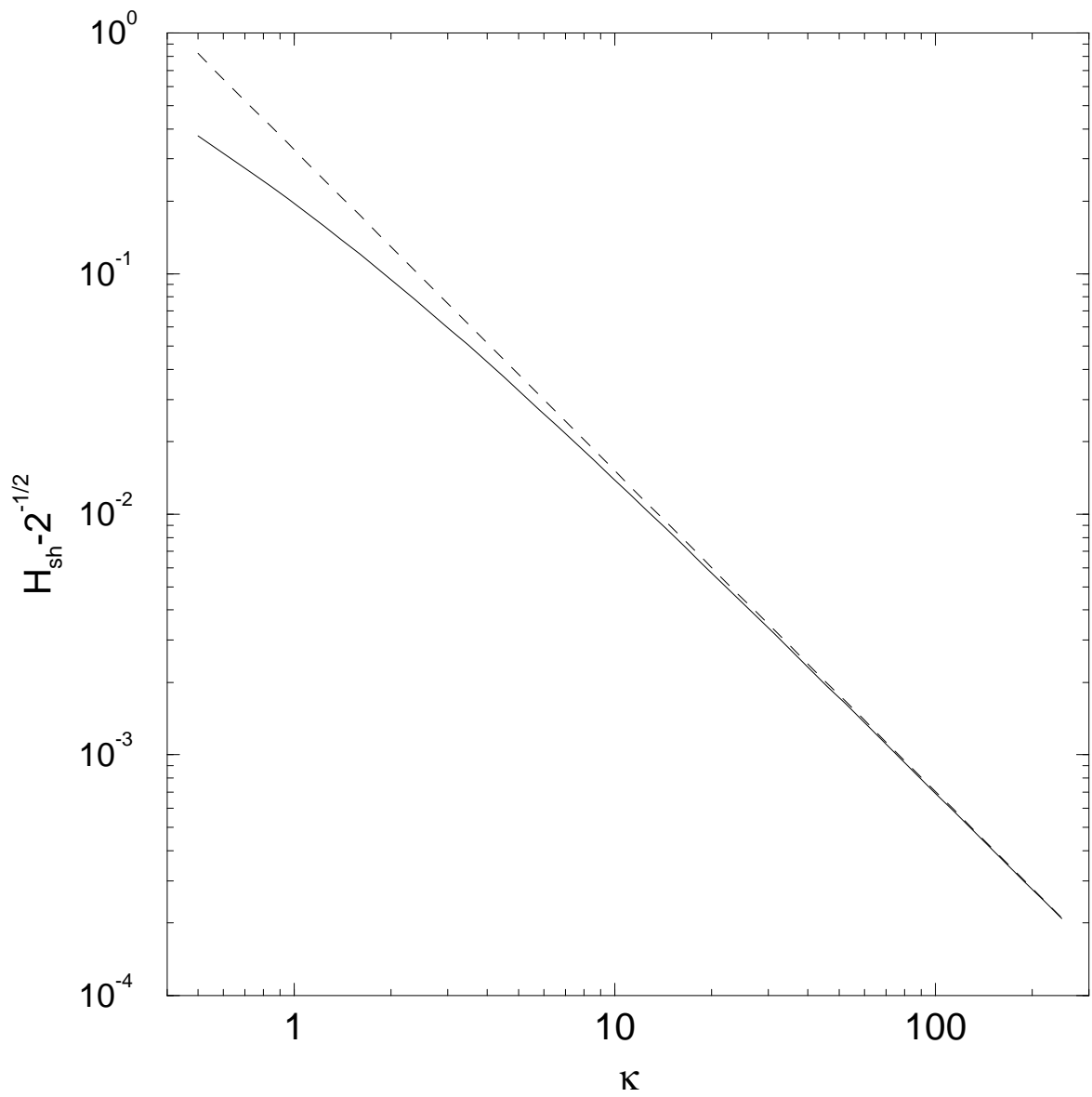
\end{document}